# TSN-IoT: A Two-Stage NOMA-Enabled Framework for Prioritized Traffic Handling in Dense IoT Networks

Shama Siddiqui[1], Anwar Ahmed Khan[2], Nicola Marchetti[3]
[1]DHA Suffa University, Karachi, Pakistan.
[2]Millennium Institute of Technology and Entrepreneurship, Karachi, Pakistan.
[3]Trinity College Dublin, Ireland.
Corresponding email: yrawna@yahoo.com

*Abstract*—With the growing applications of the Internet of Things (IoT), a major challenge is to ensure continuous connectivity while providing prioritized access. In dense IoT scenarios, synchronization may be disrupted either by the movement of nodes away from base stations or by the unavailability of reliable Global Navigation Satellite System (GNSS) signals, which can be affected by physical obstructions, multipath fading, or environmental interference, such as such as walls, buildings, moving objects, or electromagnetic noise from surrounding devices. In such contexts, distributed synchronization through Non-Orthogonal Multiple Access (NOMA) offers a promising solution, as it enables simultaneous transmission to multiple users with different power levels, supporting efficient synchronization while minimizing the signaling overhead. Moreover, NOMA also plays a vital role for dynamic priority management in dense and heterogeneous IoT environments. In this article, we proposed a Two-Stage NOMA-Enabled Framework *'TSN-IoT'* that integrates the mechanisms of conventional Precision Time Protocol (PTP) based synchronization, distributed synchronization and data transmission. The framework is designed as a four-tier architecture that facilitates prioritized data delivery from sensor nodes to the central base station. We demonstrated the performance of *'TSN-IoT'* through a healthcare use case, where intermittent connectivity and varying data priority levels present key challenges for reliable communication. Synchronization speed and end-to-end delay were evaluated through a series of simulations implemented in Python. Results show that, compared to priority-based Orthogonal Frequency Division Multiple Access (OFDMA), *TSN-IoT* achieves significantly better performance by offering improved synchronization opportunities and enabling parallel transmissions over the same sub-carrier.

## I. Introduction

IN the 6G era, IoT applications will extend far beyond traditional remote monitoring and automation. There will be hundreds of sensor nodes deployed in very close proximity, illustrating the notion of truly "dense" IoT; most deployments are expected to need efficient multiplexing techniques for meeting the demands of delay and reliability [1]. Not only will device count rise, but deployments will extend to remote areas and nodes will gain greater mobility, making it more likely that some sensors fall outside their parent gateway's range and suffer synchronization gaps. Moreover, network traffic will be heterogeneous, ranging from low-priority sensor readings (like temperature and humidity measurements) to high-priority alarms (such as medical alerts or security breaches), requiring explicit priority levels. An example of diverse IoT environment has been illustrated in Figure 1, showcasing smart healthcare, smart industry and smart vehicular networks. In such scenarios, novel approaches that jointly address multi-user access, clock alignment, and priority-aware transmission will be crucial for meeting stringent performance requirements.

Non-Orthogonal Multiple Access (NOMA) has recently emerged as a powerful tool for efficient resource sharing, enabling simultaneous uplink transmissions from multiple nodes on the same sub-band. By superimposing signals in the power domain and using Successive Interference Cancellation (SIC) at the receiver, NOMA significantly enhances spectral efficiency and reduces the access delay compared to orthogonal schemes. Given its considerable advantage for dense networks, NOMA has been widely studied for IoT deployments [2]. Moreover, when it comes to prioritized traffic, NOMA facilitates efficient spectrum utilization by eliminating the need for large bandwidth allocation to each device; instead, varying power levels are used to transmit data streams with different Quality of Service (QoS) requirements or priority levels. Additionally, Power Division Multiplexing (PDM) reduces control packet overhead, as nodes only need to occasionally negotiate power levels, unlike in Orthogonal Multiple Access (OMA) systems, where frequent coordination is often required for scheduling or contention-based channel access.

Distributed Synchronization has been proposed for dense wireless networks where there is a high probability of Global Positioning System (GPS) not being accessible by all the deployments [3]. In such systems, the conventional master slave synchronization model often fails, as it relies on every node maintaining a reliable link to a single master clock, which in dense or obstructed environments can be disrupted by backhaul delays, link asymmetries, or coverage gaps. A GPS-independent distributed synchronization scheme has been developed in [4], for collecting and updating synchronization information from neighbour nodes, while incurring a minimal signalling overhead. The iterative timing update rule converges quickly, ensuring network-wide synchronization even when number of nodes reaches hundreds. Hence, distributed synchronization promises to achieve accurate synchronization for dense networks.

A hybrid approach combining distributed synchronization and NOMA for ultra-dense wireless networks, was proposed in [5], aiming to reduce the delay associated with exchanging synchronization information among neighbouring nodes. The scheme targets scenarios where traditional GPS-based or centralized synchronization methods are impractical due to the absence of nearby base stations or limited access to external time references. To enable timely and efficient timing updates, the proposed framework integrates an uplink NOMA scheme that allows a single transmitter to communicate simultaneously with two receivers over a shared sub-band using power-domain

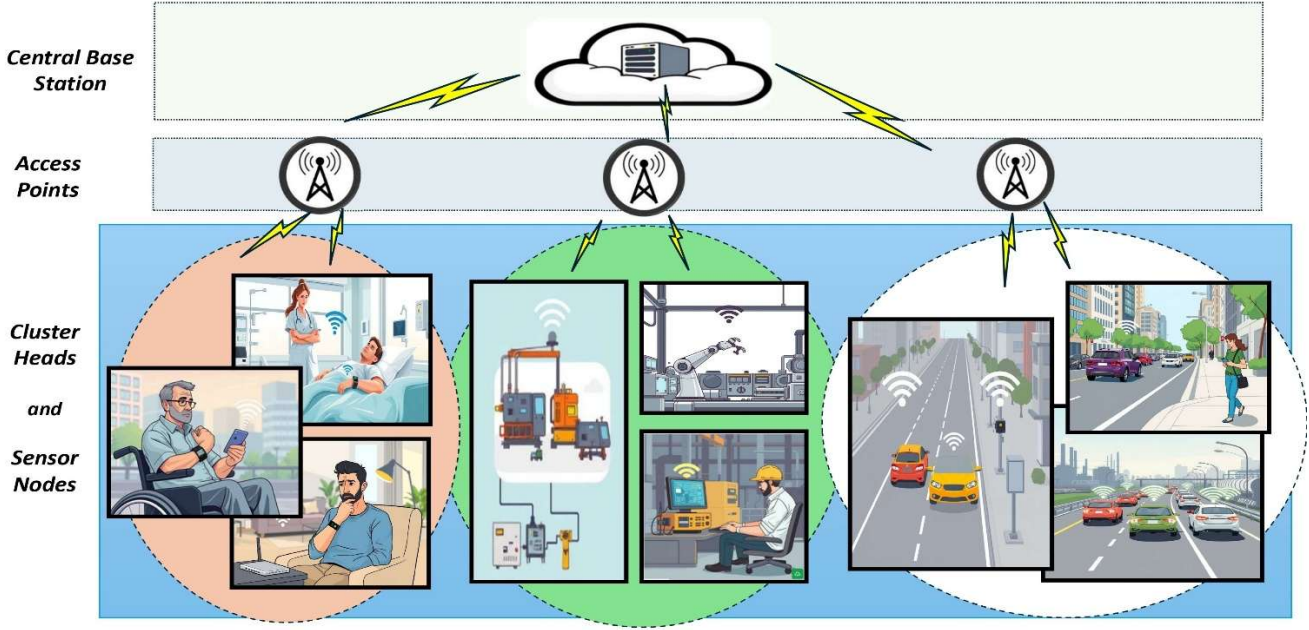

Fig. 1. Heterogeneous IoT Environment illustrating a multi-tier architecture where sensor nodes send data to their Cluster Heads (CHs), and subsequently to the Access Points (APs) and Central Base Station (CBS). Three use cases with three scenarios each are being shown: (1) Healthcare IoT with a hospitalized patient, a wheel-chair bound person in the outdoor environment, and a person at home; (2) Industrial IoT with a robotic arm, automated communication between equipment, and a user monitoring the equipment; (3) Vehicular IoT with vehicles communicating with each other, with the Roadside Unit (RSU), and with the pedestrians

multiplexing. A stable marriage-based sub-band allocation algorithm is employed to pair small cell base stations (SBSs) with frequency resources, taking into account peer effects and co-channel interference. By explicitly modelling information exchange delays and optimizing sub-band assignments and power allocation, the framework achieves significant improvements in synchronization speed and scalability while maintaining low computational complexity. These features make it suitable for deployment in large-scale, beyond-5G and 6G networks where centralized coordination is infeasible.

Although numerous frameworks have been proposed in the recent past focusing on synchronization of dense wireless networks, NOMA-enabled transmissions, and prioritized access, to the best of our knowledge, there is no unified architecture that fits specifically for IoT. For example, distributed synchronization has been integrated with NOMA in [5], but only single tier (Small Base Stations to the Mobile Core Network) is considered; there is no inclusion of lower layers of IoT architecture such as Sensor Nodes (SNs) or Cluster Heads (CHs); moreover, the model only focuses on distributed synchronization phase, and does not address the actual data transmission process. Similarly, NOMA has been used for prioritized data transmission for IoT networks in [6], but the work only incorporates beacon transmission for achieving synchronization with the base station. Therefore, we identified a gap to develop an integrated multi-tier model for dense IoT networks.

This paper focuses on a two-stage distributed synchronization and data transmission framework, named TSN-IoT, developed using semi-grant free uplink NOMA, for facilitating efficient transmission of prioritized traffic in dense IoT environments. In semi-grant free uplink NOMA, some users are scheduled through grant-based access while others transmit without explicit grants, thereby combining the reliability of controlled access with the low-latency benefits of grant free access. Unlike the previous work, we use NOMA both for synchronization and data transmission, enabling the data transmission from Sensor Nodes (SNs) to the Base Station (BS), via Cluster Heads (CHs) and Access Points (APs). We discuss the details of sub-band allocation for each communication tier, the use of Precision Time Protocol (PTP) and distributed synchronization along with the data transmission mechanism. We develop a use case of hospital environment for demonstrating the performance of TSN-IoT in terms of delay and synchronization speed. Moreover, we discuss the opportunities, challenges and open directions for developing hybrid communication frameworks for next generation, dense multi-tier IoT architectures. The contributions of this work are as follows:

- We propose a hybrid framework, named *TSN-IoT*, that employs NOMA for integrating distributed synchronization and prioritized data transmission in dense IoT networks.
- We integrate and implement protocols and sub-band allocation schemes within the TSN-IoT framework.
- We evaluate TSN-IoT through a healthcare use case focusing on delay performance and priority management through simulations.

## II. TECHNICAL BACKGROUND

In this section, we review the three key technologies that underpin our TSN-IoT framework. We begin with the principles and benefits of NOMA, which supports simultaneous multiuser access in dense deployments and has also been applied in IoT scenarios to facilitate prioritized data transmission. Next, we discuss GPS-independent, peer-to-peer synchronization techniques that maintain tight clock alignment without relying on a central master clock, with a focus on their application in IoT environments. Finally, we cover priority aware transmission schemes designed to handle heterogeneous IoT traffic and ensure timely delivery of critical data.

*A. NOMA for Multi-user Access in IoT*

Recent developments in NOMA have focused on enhancing its adaptability to diverse IoT communication needs by refining how spectrum is shared among users. One such approach, as explored in recent work on sub-band Multi-Carrier NOMA (MC-NOMA), leverages the division of available bandwidth into multiple sub-bands, each serving multiple users through power-domain multiplexing [7]; this setup allows for fine grained pairing of users with complementary channel gains, enhancing throughput while keeping interference manageable through SIC. In addition to improved capacity, such schemes offer adaptability, enabling devices with different traffic demands to share spectrum more effectively, which is an essential feature in IoT scenarios where low-latency and high-reliability traffic may coexist with delay-tolerant background data.

Beyond static sub-band allocation, recent studies have explored how NOMA can be integrated into random access schemes to improve performance under sporadic or uncoordinated transmissions, which are typical in massive IoT deployments [8]. The use of grant-free NOMA, where devices transmit without prior scheduling, enables faster channel access and reduces signalling overhead, especially when coupled with efficient user activity detection techniques. This is particularly beneficial in dense scenarios where coordinated scheduling may not scale well. Advanced approaches such as contention resolution diversity slotted ALOHA (CRDSA) combined with NOMA have shown promise in resolving collisions while maintaining support for simultaneous multi-user transmissions.

*B. Evolution of Clock Sync: From Master Clocks to Distributed Algorithms*

The progression of wireless network design has led to a fundamental transformation in clock synchronization mechanisms: from rigid, master-slave architectures toward more flexible, distributed algorithms. In traditional models, synchronization relies on a single master node broadcasting timing updates to all others, assuming consistent connectivity and symmetric delays. However, as networks become denser, more mobile, and heterogeneous, such assumptions break down. Distributed synchronization techniques instead, enable each node to collaborate locally with its neighbours to gradually align their clocks. By removing the dependence on a centralized authority, such algorithms allow the network to adapt dynamically to changes in topology, mobility, or failure conditions, making them particularly well-suited for large-scale IoT environments.

To address the limitations of centralized synchronization in dense IoT settings, clustering-based distributed synchronization has emerged as a practical solution. Instead of relying on a global master clock, nodes are organized into logical clusters, each governed by a designated cluster head responsible for coordinating timing updates. Recent strategies enhance the cluster-based approach by intelligently selecting cluster heads based on signal quality and link stability, ensuring reliable propagation of timing information across the network [9]. Once intra-cluster synchronization is achieved, inter-cluster alignment is maintained through periodic exchanges between heads, enabling scalable and resilient timing across the entire deployment. Such localized, multi-level synchronization schemes provide a middle ground between fully decentralized algorithms and rigid centralized protocols, offering improved performance in industrial and mission-critical IoT scenarios.

*C. Priority-Aware Communication in IoT*

In high-density IoT environments such as health monitoring systems, communication networks must account for the varying criticality of data generated by devices; various approaches have been proposed to address this requirement. For example, message-passing-based scheduling algorithms have been developed that assign transmission priorities based on delay tolerance and service urgency. A model developed in [10] represents the system as a priority graph where each device communicates its data requirements to a central decision unit, which then allocates resources using iterative optimization. The message passing framework, therefore, allows critical healthcare data, such as emergency alarms or vital sign deviations, to be transmitted with minimal latency.

An Orthogonal Frequency Division Multiple Access (OFDMA) scheduler has also been proposed for managing prioritized healthcare data in [11]; this scheme dynamically prioritizes stations based on the sizes of their queues for different Access Categories (ACs) and their recent transmission history. By leveraging the multi-user resource unit (RU) allocation capability of Wi-Fi 7, the scheduler assigns higher transmission opportunities to ACs carrying urgent medical data, while background traffic is deferred to later slots. It has been observed that OFDMA scheduler achieves substantial reductions in latency for critical health packets, while maintaining acceptable overall system efficiency.

NOMA-based approaches have increasingly been employed to handle data prioritization challenges in massive IoT deployments where urgent messages must coexist with routine data streams. In [12], a hybrid user-grouping mechanism is presented, combining both power-domain and code-domain NOMA to accommodate heterogeneous devices with varying priority levels. By dynamically assigning users to groups based on channel quality and urgency of data, the system ensures efficient superposition coding and robust decoding at the receiver end. Simulation results show that this hybrid method significantly improves reliability and spectral efficiency in mixed-priority traffic scenarios. Moreover, the scalable grant free framework proposed in [13] incorporates code-domain NOMA with multi-signature spreading to support low-latency access for a massive number of IoT devices.

## III. TSN-IoT: An Integrated Synchronization and Data Transmission Framework

To address the limitations of existing synchronization and transmission models in dense IoT environments, we propose TSN-IoT, a unified framework that leverages uplink NOMA across multiple tiers. This section presents the architecture and operational workflow of TSN-IoT, highlighting how it integrates distributed synchronization with priority-aware data transmission.

### A. Architecture and Functional Layers

The TSN-IoT architecture consists of four hierarchical layers: SNs, CHs, APs, and a CBS, working together to enable synchronized, priority-aware communication across the network; this architecture is illustrated in Figure 2, and maps directly to the high-level use cases shown earlier in Figure 1. At the bottom-most tier, SNs act as sensing units and communicate with their respective CHs using semi-grant free NOMA. Simultaneous transmissions from multiple SNs are superimposed in the power domain, allowing high-priority traffic (e.g., alarms or real-time events) to be transmitted at higher power levels. Each SN begins by transmitting a short info packet at a predefined power level to establish its transmission intent and share its data priority. For example, when two SNs, say SN1 and SN2, share the same sub-band, one may transmit its info packet at Power Level 1 (PL1) and the other at Power Level 2 (PL2). The CH uses signal strength and embedded metadata to compare priorities. If SN1 has higher or equal priority than SN2, SN1 transmits the actual data at PL1, while SN2 uses PL2, where we assume PL1 > PL2. Otherwise, the assignment is reversed. The CH then applies SIC to decode the superimposed signals, always prioritizing high-priority data for early decoding to ensure QoS.

CHs operate at the second tier, and serve both the synchronization and forwarding roles. They establish synchronization with their associated APs using PTP, while also coordinating distributed synchronization with the peer CHs using a weighted consensus approach, if needed. This two-phase timing model ensures that even in the absence of GPS coverage or stable backhaul links, local synchronization can be preserved with minimal signaling overhead. CHs also aggregate incoming data and relay it, via uplink NOMA to the APs.

The third tier comprises APs, which collect and forward aggregated traffic from multiple CHs. Each AP synchronizes with the CBS using PTP and also serves as a reference node for downstream CHs. APs communicate with CHs using NOMA based transmission, with sub-band allocation optimized to minimize co-channel interference while respecting traffic priority levels.

At the highest tier, the CBS serves as the global timing anchor and data sink. It propagates time synchronization messages to all APs and can initiate resource reallocation or intervention when congestion or high-priority traffic is detected. While the CBS initiates primary synchronization, the lower tiers also maintain synchronization through periodic distributed synchronization cycles ensuring continuity in case of temporary disconnections or topology changes.

### B. Synchronization and Data Transmission Workflow

The operational flow of TSN-IoT is presented in Figure 3. The operation begins with initial synchronization, where the CBS propagates timing information to APs using PTP. The APs then synchronize CHs, which further synchronize SNs. If any node fails to synchronize during this phase, the framework transitions to distributed synchronization, initiated first between APs and then among CHs using a consensus-based update scheme.

Data transmission is performed in parallel using uplink NOMA across all tiers. SNs transmit to CHs, CHs forward to APs, and APs relay to the CBS. If any node is temporarily disconnected from its upstream tier, the data is forwarded via neighbouring nodes at the same level, i.e., CH-to-CH or AP-to-AP (referred as limited forwarding), to ensure timely delivery. This fallback mechanism enhances network resilience and ensures that data, particularly from high-priority applications, continues to flow even in partially connected deployments.

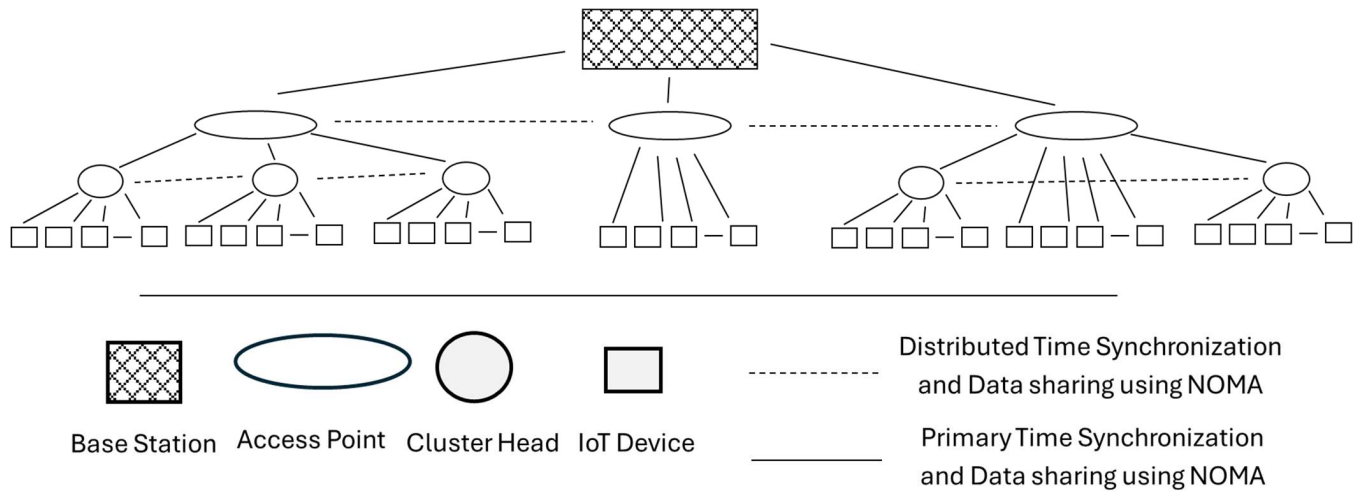

Fig. 2. System architecture of TSN-IoT. Three example configurations are illustrated: the first and third depict a four-tier setup, while the second shows a three-tier setup, as in some scenarios, a single component may serve as CH and AP, for instance a smart phone

## IV. SIMULATION STUDY: TSN-IoT IN HEALTHCARE IoT

We evaluated TSN-IoT in a dense healthcare IoT scenario, with the number of SNs varied from 400 to 1600, while the number of CHs and APs remained fixed at 40 and 4, respectively, and a single CBS acting as the grandmaster clock. These values reflect the characteristics of hospital environments, where hundreds of patients and medical devices generate continuous streams of heterogeneous data. High SN counts emulate large deployments of wearable sensors, infusion pumps, and monitoring devices in wards and ICUs, while urgent versus normal traffic models correspond to time-critical alerts (e.g., arrhythmia or oxygen desaturation) versus routine vital signs. Synchronization cycles were set to 10 minutes, with PTP used in the initial phase and distributed consensus-based synchronization between CHs and APs triggered only when PTP failed. The choice of a 10-minute cycle is consistent with periodic medical data logging and time-sensitive sensor fusion requirements reported in healthcare IoT studies, and similar large-scale, latency-sensitive sensing configurations have been documented in hospital IoT deployments and smart healthcare monitoring frameworks [14].

Two traffic categories, urgent and normal, were generated at different rates and transmitted at different power levels, with NOMA pairing applied in all tiers. Frequency sub-band reuse, Signal-to-Interference-plus-Noise Ratio (SINR) and power thresholds were enforced, and Rayleigh fading was applied to model realistic wireless channel conditions. These thresholds were used to enforce realistic reception and decoding conditions in the simulation. Specifically, the power threshold ensured that received signals below a minimum strength were treated as unsuccessful transmissions, while the SINR threshold determined whether SIC in NOMA decoding could be applied successfully.

Data rates were configured separately for each tier to reflect realistic traffic loads, and data packets from SNs were aggregated at CHs before transmission, while APs forwarded CH data without further aggregation. The tier-specific data rates were chosen to reflect the relative traffic volumes generated at different levels of the hierarchy in dense healthcare IoT settings. At the SN–CH tier, high-frequency sensing generates large volumes of small packets, which are aggregated by CHs before transmission. At the CH–AP and AP–CBS tiers, the traffic consists of aggregated flows with correspondingly higher data volumes but lower packet generation frequency, consistent with prior studies on hierarchical IoT architectures [15].

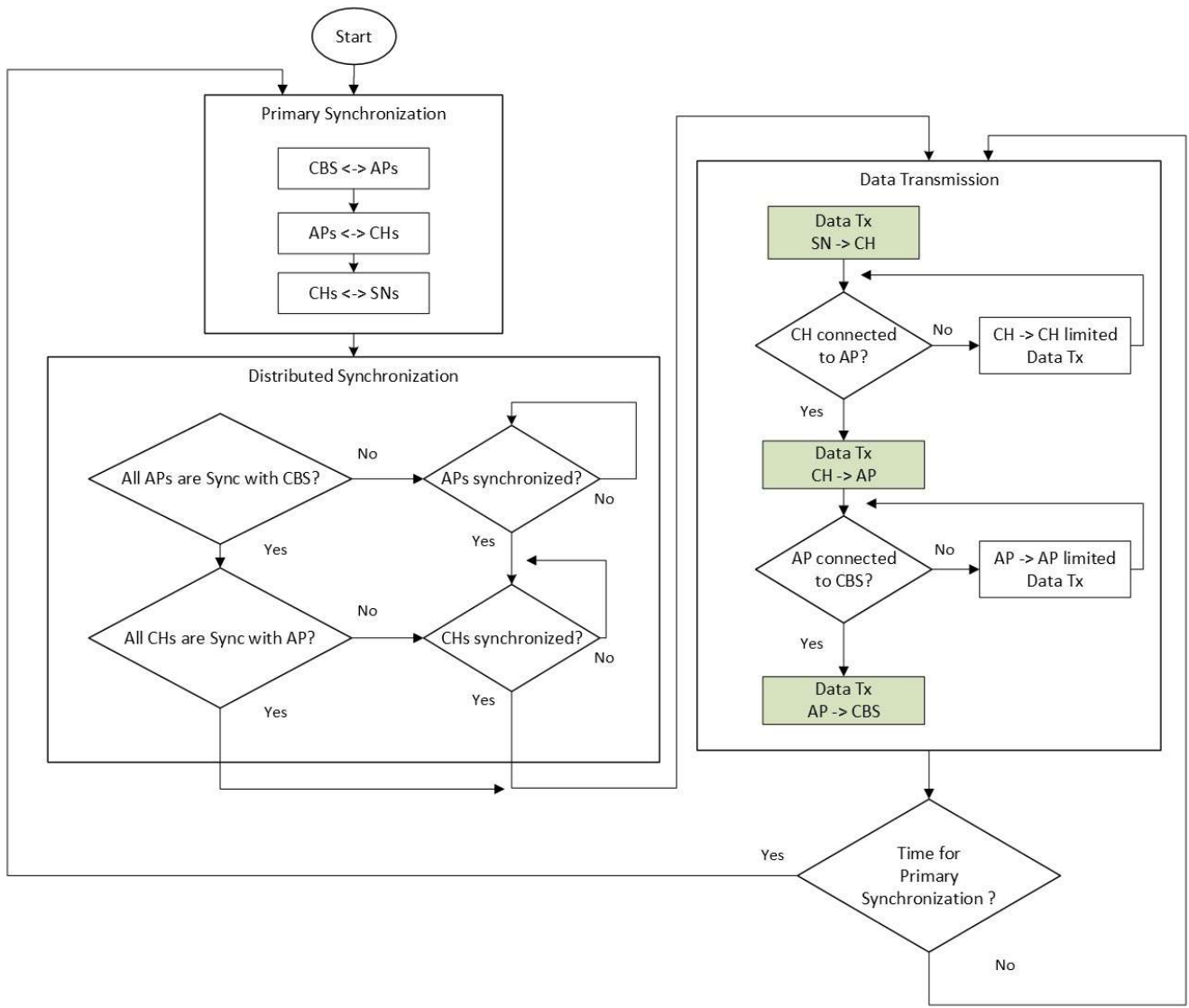

Fig. 3. Flow diagram of TSN-IoT showing periodic primary synchronization, conditional distributed synchronization, and data transfer from SNs to the CBS via CHs and APs.

The SIC threshold of 3 dB was used as the minimum SINR difference required for SIC cancellation in NOMA decoding. NOMA user pairs were restricted to two nodes per superposition, meaning that at each tier (SN–CH, CH–AP, AP–CBS) a maximum of two users shared the same resource block for SIC. A CH aggregation factor of 0.6 was used, implying that 60% of the aggregated packet size corresponded to payload data after header overheads were removed, and this was applied whenever multiple SN packets were combined at a CH. The reliability factor of 0.9 was defined in the simulation as a probabilistic success rate applied uniformly to synchronization, decoding, and transmission processes. In practice, this meant that each synchronization attempt, each NOMA-based SIC operation, and each data packet transmission succeeded with a probability of 0.9 and failed otherwise. The value of 0.9 was selected in line with common reliability modelling assumptions in wireless communication studies, where sub perfect success rates (typically between 0.85 and 0.95) are used to represent realistic operating conditions. The simulation parameters have been listed in Table I.

Figure 4 and Figure 5 present the performance comparison between TSN-IoT and a priority-based OFDMA scheduler under the simulated healthcare IoT setting. The results are plotted by varying the number of sensor nodes while observing two key performance indicators, synchronization time and end to-end delay, highlighting how each framework scales with increasing network density. In Figure 4, synchronization time is plotted against the number of sensor nodes. While both schemes show increasing synchronization time with growing network density, TSN-IoT consistently achieves much lower values than OFDMA. For example, at 400 nodes, TSN-IoT requires only about 0.8 ms, compared to nearly 3 ms for OFDMA, and at 1600 nodes, TSN-IoT remains below 3 ms while OFDMA exceeds 12 ms. This performance gain is attributed to the dual synchronization strategy of TSN-IoT, which combines periodic PTP-based primary synchronization with distributed consensus updates among CHs and APs when required, ensuring robustness even under dense deployments. Furthermore, TSN-IoT leverages NOMA, allowing multiple nodes to transmit simultaneously on the same subcarrier, thereby avoiding the contention and scheduling delays inherent in OFDMA. In contrast, the OFDMA baseline relies solely on conventional synchronization, making it more susceptible to signaling delays and congestion.

Furthermore, in Figure 5, the average end-to-end delay is shown for two sub-band allocation strategies at the CH level. Each CH provided 5 sub-bands for 10 sensor nodes, with the first scenario reserving 4 sub-bands for urgent traffic and 1 for normal, while the second reserved 3 for urgent and 2 for normal. As expected, the configuration with 4 sub-bands dedicated to urgent traffic achieved the lowest delay for high-priority packets, since urgent data streams encountered minimal contention. This result highlights the effectiveness of prioritization in sub-band allocation, where increasing the share for urgent traffic directly reduces latency for critical data. Additionally, TSN-IoT employs an info packet mechanism in combination with NOMA, enabling nodes to declare priority before transmission. This, coupled with simultaneous transmission on the same subcarrier and SIC decoding, ensures that urgent packets are consistently delivered with reduced delay compared to the OFDMA baseline. When the number of nodes increases, however, the total number of sub-bands remains fixed, leading to higher contention and consequently increased delay.

TABLE I KEY SIMULATION PARAMETERS

| Parameter | Value | Notes |
|---|---|---|
| SN count sweep | 400 → 1600 (step 200) | Derived from 40 CHs; 10–40 SNs/CH |
| CH / AP / CBS | 40 / 4 / 1 | Fixed topology |
| Hop distances | CH–SN: 15 m; AP–CH: 20 m; CBS–AP: 30 m | Used for path loss & PTP timing |
| Channel data rates | SN→CH: 1 Mb/s; CH→AP: 10 Mb/s; AP→CBS: 100 Mb/s | Tier-specific rates |
| NOMA users per pair | 2 (all tiers) | Superposition size |
| Parallel sub-bands (total) | CBS: 2; AP: 5; CH: 5 | Urgent/Normal sub-band split: (1/1), (4/1) or (3/2), (4/1) or (3/2) |
| Max retries | 3 | For sync & data attempts |
| PTP messages & size | 3 × 4 B; proc. delay 10 $\mu$s each | Total bits = 96; includes propagation delay |
| Distributed sync | Slot 10 $\mu$s; 3–20 iterations | Consensus-based timing update |
| Base SNR & density penalty | 10 dB base; −0.25 dB per +40 SN | Effective SNR vs. density |
| Path-loss exponent | 2.5 | For large-scale fading |
| SIC threshold (NOMA) | 3 dB | Decoding condition |
| Packet size | 128 B (8 B header, 120 B payload) | Per packet |
| Per-SN traffic (per min) | Normal: 10 pkts; Urgent: 10 pkts | Traffic load |
| CH aggregation factor | 0.6 (payload) | Applied when SNS per CH > 1 |
| Averaging | 20,000 minutes | For stable means |

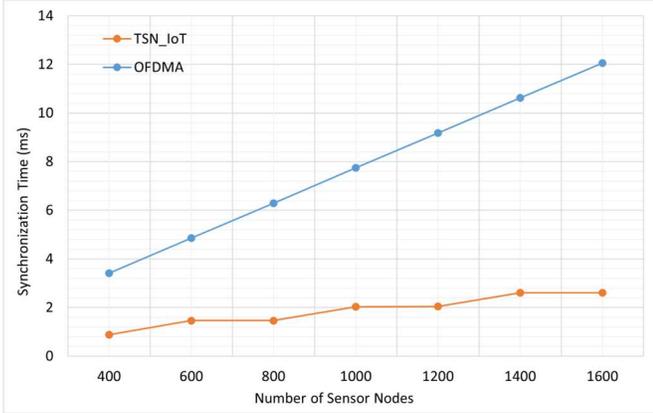

Fig. 4. Synchronization time comparison of TSN-IoT and priority-based OFDMA scheduler under varying numbers of sensor nodes, obtained from Python simulations.

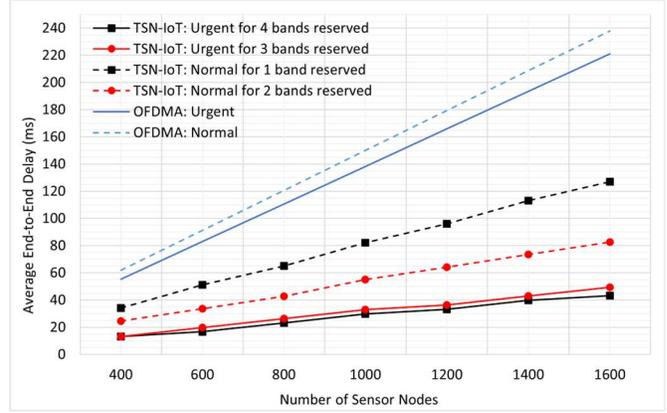

Fig. 5. End-to-end delay comparison for TSN-IoT and priority-based OFDMA scheduler. Impact of reserving certain bands specifically for urgent traffic has also been studied.

## V. Toward Real-World Adoption: Challenges and Prospects

Moving from theory to deployment, the *TSN-IoT* framework faces practical challenges across access, synchronization, and priority handling. The following subsections highlight these issues and outline potential solutions for real-world adoption.

### A. NOMA-Based Access in Practice

While NOMA offers substantial gains in spectral efficiency and priority handling, its implementation in large-scale IoT networks presents several challenges. SIC incurs computational overhead and can be sensitive to power allocation errors, especially when devices operate under low-cost hardware constraints typical of IoT deployments. Imperfect channel state information (CSI) further complicates power-domain separation, leading to increased error rates and degraded quality of service. Although in theory NOMA can support $n>2$ distinct power levels within a sub-band, in practice most implementations are limited to only two power levels, restricting the degree of multiplexing achievable.

Prospective solutions include lightweight SIC algorithms optimized for resource-constrained devices, adaptive power allocation strategies that incorporate fairness metrics, and hybrid NOMA–OFDMA approaches where orthogonal resources are dynamically assigned to overloaded tiers. These directions can bridge the gap between NOMA's theoretical benefits and its practical feasibility in dense IoT ecosystems.

### B. Limitations of Distributed Synchronization

Distributed peer-to-peer synchronization reduces dependency on GPS or a single master clock, but it is not without challenges in dense IoT deployments. Iterative consensus mechanisms rely on repeated message exchanges, which may introduce significant overhead and additional delay as the number of nodes scales. Link asymmetries, packet loss, and fluctuating wireless conditions (e.g., due to fading or interference) can further degrade synchronization accuracy, especially when updates are propagated over multiple hops.

Future prospects include the design of lightweight consensus algorithms that minimize message exchanges, probabilistic synchronization methods that tolerate occasional packet losses, and cross-layer optimization techniques where synchronization is jointly managed with data transmission. Incorporating machine learning to predict drift patterns and correct timing deviations in advance may also enhance robustness in real-world deployments.

### C. Priority-Aware Transmission Under Constraints

While priority-aware mechanisms ensure that urgent traffic is delivered ahead of delay-tolerant data, practical limitations arise when applying them in dense IoT networks. The number of available sub-bands is fixed, so as the number of nodes grows, multiple devices must contend for the same resources, which can still result in collisions or queuing delays. Allocating too many resources to urgent traffic may also lead to starvation of normal flows, undermining fairness. Furthermore, lightweight devices often lack the processing capability to support complex scheduling or classification algorithms.

Potential solutions include adaptive sub-band reservation schemes that dynamically balance urgent and normal traffic based on real-time load, as well as reinforcement learning methods that optimize priority allocation under changing network conditions. Edge-assisted scheduling could further relieve end devices from heavy computation while preserving fairness and low-latency delivery.

## VI. Conclusion

In this paper, we introduced TSN-IoT, a two-stage NOMA enabled framework that integrates distributed synchronization with priority-aware data transmission for dense IoT environments. Unlike conventional approaches that treat synchronization and data delivery separately, TSN-IoT unifies these functions across multiple tiers, from sensor nodes to the central base station, ensuring robust timing alignment and efficient spectrum utilization under heterogeneous traffic demands.

Simulation results in a healthcare IoT setting demonstrated that TSN-IoT consistently outperforms a priority-based OFDMA scheduler, achieving sub-millisecond synchronization time and significantly reducing end-to-end delay even as the number of sensor nodes scaled into the thousands. These improvements were attributed to NOMA's ability to superimpose transmissions on the same subcarrier, the use of distributed consensus for resilience when primary synchronization fails, and the info packet mechanism that enables explicit priority handling.

Looking ahead, future work will extend TSN-IoT by incorporating additional performance metrics such as throughput and fairness, evaluating its scalability across diverse IoT domains, and developing lightweight optimization models to refine power allocation and sub-band assignment. Machine learning will also be explored to enhance adaptability, enabling the framework to predict traffic patterns, optimize synchronization overhead, and dynamically adjust priority levels. By addressing these directions, TSN-IoT can serve as a practical enabler for next-generation IoT systems, supporting real-time, reliable, and priority-sensitive communication in ultra-dense deployments.

## Acknowledgment

This contribution is supported by HORIZON-MSCA2022-SE-01-01 project COALESCE under Grant Number 10113073.

## VII. Biography Section

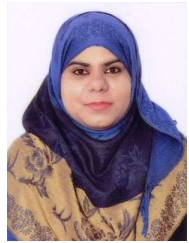

Shama Siddiqui did M.Sc in Applied Physics (with specialization in Electronics) from University of Karachi, Pakistan in 2006. She did her MS (Computer Science) from Iqra University, Karachi, Pakistan in 2010. She did PhD from Faculty of Computer Science, Institute of Business Administration (IBA) Karachi in 2018. She worked as a lecturer computer Science at Institute of Business Management (IoBM) Karachi and as adjunct faculty at IBA. At present, she is associated with DHA Suffa University, Karachi as an Associate Professor at the department of Computer Science. She is a visiting researcher at Trinity College Dublin, Ireland under EU-Horizon MCSA program. Her research interests include performance analysis of MAC protocols for Next Generation 6G networks, developing performance analysis frameworks and remote health monitoring applications of IoT. She has published over 50 articles in internationally reputed journals and conferences. She has active international collaborations with various organizations including South East Technological University Ireland, University of Glasgow, King Abdul Aziz University, and University of Missouri at Kansas City.

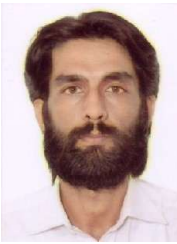

Anwar Ahmed Khan did B.E (Electronics) from Hamdard University Karachi, Pakistan in 2004, M.Sc in Applied Physics (with specialization in Electronics) from University of Karachi in 2006, M.E (Electronics) from NED University of Engineering & Technology, Karachi in 2015 and PhD from Faculty of Computer Science, Institute of Business Administration (IBA) Karachi in 2020. He worked at Hamdard University as a lab instructor, at Oil & Gas Development Corporation Ltd. as a trainee instrumentation engineer, at National Institute of Electronics as a design engineer and as adjunct faculty at IBA, at Sindh Institute of Management and Technology (SIMT) Karachi as assistant professor and Head of department for computer science. At present, Dr. Khan works at Millennium Institute of Technology and Entrepreneurship as Associate Professor in the Computing Department. Dr. Khan is actively engaged in research in the areas of WSN and IoT and has published and presented his work at reputable venues such as IEEE ICC, IEEE COMPSAC, Mobicom, China-Com, IEEE Sensors Journal and Wireless Personal Communications. Furthermore, he also has experience of organizing conferences such as IEEE ICICT, which is one of the top conferences held at Pakistan, arranged by IBA, Karachi. He has been invited at various universities for delivering guest lectures, particularly focused on testbed development and demonstrations of Internet of Things.

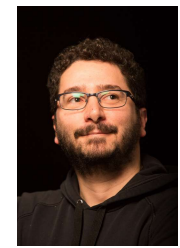

Nicola Marchetti is Professor in Wireless Communications at Trinity College Dublin, Ireland. He is an IEEE Senior Member, a Fellow of Trinity College, and was an IEEE Communications Society Distinguished Lecturer. He received the PhD in Wireless Communications from Aalborg University, Denmark in 2007, the MSc in Electronic Engineering from University of Ferrara, Italy in 2003, and the MSc in Mathematics from Aalborg University in 2010. He has authored more than 200 journals and conference papers, 2 books and 9 book chapters, holds 4 patents, and received 4 best paper awards. His research interests span Classical and Quantum Networks, Communications and AI for Environmental Monitoring, Complex Systems, and Mathematics for Communication & Computation. He serves as Technical Editor for IEEE Wireless Communications and has served as an Associate Editor for IEEE Network, the IEEE Internet of Things Journal and the EURASIP Journal on Wireless Communications and Networking. He has an extensive experience giving research lectures, having delivered to date 6 keynotes, 9 IEEE ComSoc distinguished lectures, 20 invited talks, 8 tutorials at international conferences, and 8 international PhD courses.